\documentclass[conference]{IEEEtran}
\IEEEoverridecommandlockouts
\usepackage{cite}
\usepackage{amsmath,amssymb,amsfonts}
\usepackage{algorithmic}
\usepackage{graphicx}
\usepackage{textcomp}
\usepackage{xcolor}

\usepackage{url}
\usepackage{subcaption} 

\def\BibTeX{{\rm B\kern-.05em{\sc i\kern-.025em b}\kern-.08em
    T\kern-.1667em\lower.7ex\hbox{E}\kern-.125emX}}
\begin{document}

\title{
Zema Dataset: A Comprehensive Study of Yaredawi Zema with a Focus on Horologium Chants 
}

\author{
    \IEEEauthorblockN{Mequanent Argaw Muluneh$^{1,3,4}$ \and
    Yan-Tsung Peng$^3$ \and
    Worku Abebe Degife$^2$ \and
    Nigussie Abate Tadesse$^5$ \and
    Aknachew Mebreku Demeku$^6$ \hspace{1cm} Li Su$^1$ 
    }   
    \IEEEauthorblockA{$^1$ Institute of Information Science, Academia Sinica, Taipei, Taiwan} 
    \IEEEauthorblockA{$^2$ Department of Information System, University of Gondar, Gondar, Ethiopia} 
    \IEEEauthorblockA{$^3$ Department of Computer Science, National Chengchi University, Taipei, Taiwan} 
    \IEEEauthorblockA{$^4$ Social Networks and Human-Centered Computing Program, Taiwan International Graduate Program}

    \IEEEauthorblockA{$^5$ Department of Information Management, National Taiwan University of Science and Technology, Taipei, Taiwan} 
    \IEEEauthorblockA{$^6$ Department of Material Science and Engineering, National Taiwan University of Science and Technology, Taipei, Taiwan}
}  

\maketitle 

\begin{abstract}
Computational music research plays a critical role in advancing music production, distribution, and understanding across various musical styles in the world. Despite the immense cultural and religious significance, the Ethiopian Orthodox Tewahedo Church (EOTC) chants are relatively underrepresented in computational music research. This paper contributes to this field by introducing a new dataset specifically tailored for the analysis of EOTC chants, also known as \emph{Yaredawi Zema}. This work provides a comprehensive overview of a 10-hour dataset, 369 instances, creation, and curation process, including rigorous quality assurance measures. Our dataset has a detailed word-level temporal boundary and reading tone annotation along with the corresponding chanting mode label of audios. Moreover, we have also identified the chanting options associated with multiple chanting notations in the manuscript by annotating them accordingly. 
Our goal in making this dataset available to the public\footnote{Dataset: \url{https://github.com/mequanent/ChantingModeClassification}} is to encourage more research and study of EOTC chants, including lyrics transcription, lyric-to-audio alignment, and music generation tasks. Such research work will advance knowledge and efforts to preserve this distinctive liturgical music, a priceless cultural artifact for the Ethiopian people. 

\end{abstract}





\section{Introduction} \label{sec:introuduction}
The teaching-learning process in the spiritual schools of the Ethiopian Orthodox Tewahedo Church (EOTC) takes a long time. Due to the pleasing nature of the EOTC chants, also known as \emph{Yaredawi Zema}, students devote their lives to learning various categories of chants. 
\emph{Digua}, the largest book of chant, used to take 40 years to master with oral training \cite{berhanu-akal:2012}. This oral training
, which takes a lot of time, 
was supported 
only 
by handwriting on parchments made from goat skin \cite{Kebede:80}. By the introduction of printed books with notations, the minimum number of years required to master \emph{Digua} reduced significantly to 8 years 
if excluding the years for pre-requisite steps 
\cite{mezmur-tsegaye-tourist:2011}. 
Moreover, recording technologies supported by mobile applications are also helping a lot to facilitate self-learning. 
In recent years, 
advancements in artificial intelligence (AI) 
have facilitated various 
educational \cite{PIKHART-ai4language-2020, ZHANG-AI-Education:2021} and 
cultural 
\cite{youtube-recsys:2016, netflix-recsys:2016} applications 
of music and language
. 
Utilizing computational tools for the transcription, analysis, and digitization of music data across the world has emerged as a novel trend in ethnomusicology research,  \cite{han2023finding,rosenzweig2020erkomaishvili,nikzat2022kdc,caro2014creating,nadkarni2023exploring,lisu:amt-sitar-music}, music theory development \cite{moss2023line,lieck2020tonal}, as well as the preservation and revitalization of music cultural heritage \cite{music-culture:2021,weiss2021schubert,stepputat2019digital,rosenzweig2022computer}. 
Such 
computational music research provides powerful tools for analyzing and studying various musical styles. However, 
they 
are often focused on a limited number of musical traditions \cite{lisu:amt-sitar-music}. For underrepresented music cultures, diversifying the music datasets is also important in the aspect of avoiding cultural bias in the data \cite{holzapfel2018ethical}. 

This paper aims to fill this gap by introducing a new dataset specifically designed for computational analysis and understanding of the EOTC chants. Our initiative in developing this dataset is driven by the goal of deepening the understanding and appreciation of \emph{Yaredawi Zema}, which are a significant aspect of the spiritual and cultural heritage of Ethiopia. 
By providing the chants' audio, lyric-text, and reading tone labels, this dataset is also for multiple purposes, available 
for tasks such as chanting mode classification, lyric transcription, lyric-to-audio alignment, and generation. 
Moreover, 
the inclusion of chanting options' labels 
can
support 
further musicological 
analysis and understanding tasks 
of EOTC chants. To our knowledge, our work represents a first attempt to construct a 
dataset for computational research of \emph{Yaredawi Zema}.

The remainder of this paper is structured as follows: Section \ref{sec:background} provides an overview of the EOTC spiritual schools, chanting modes, notation, and chanting options within each mode. Section \ref{sec:dataset} details the dataset preparation process and presents a summary of its contents. Finally, Section \ref{sec:conclusion} concludes the paper with suggestions for future work on the dataset.

\section{Background}\label{sec:background} 

\subsection{Spiritual Schools}

The spiritual schools of the EOTC provide instruction in two fundamental areas that are interconnected. The first area focuses on chants that are sacred music, collectively referred to as \emph{zema} in both the Ge'ez and Amharic languages. The second area encompasses the study of the Geez language, the spiritual and philosophical poetry known as \emph{Qine}, and the exegesis of scriptures. These areas of study are taught in 
specialized schools, including the 
\emph{Nibab-bet} (Ge'ez language reading practice), \emph{Zema-bet} (basic to advanced level of one chanting category), \emph{Qidase-bet} (liturgical chants used in mass services), \emph{Qine-bet} (Ge'ez language perfection and spiritual philosophy in the form of poetry), \emph{Aquaquam-bet} (advanced chanting with accompaniments), \emph{YeMetsahift Tirguame-bet} (exegesis of scriptures). All these specialized schools use Ge'ez as the primary language for instruction, with Amharic also used, especially in scripture exegesis.

The \emph{Nibab-bet} is the foundational school where students begin their practice of reading the Ge'ez language. Ge'ez is a tonal language that requires each word to be read in one of five distinct tones: \emph{tenesh} (upper tone), \emph{seyyaf} (upper tone, different from \emph{tenesh} by ending alphabets), \emph{tetay} (smooth tone), \emph{wedaki} (longer smooth tone), and \emph{tenababi} (tone for dependent prefix words). Mastery of these tones is essential for differentiating similar chanting notations and serves as the gateway to all other fields of study. Our dataset includes the corresponding labels for the reading tone of each word as it is sung in the chants. 
For detailed explanations about each school, the readers are advised to refer to the corresponding section in \cite{mezmur-tsegaye-tourist:2011}.


Given that each EOTC spiritual school typically has only one teacher per location \cite{Dagne-eotc-org}, it is impressive how a single instructor manages to teach students at different levels of study. An effective way to handle this is to have higher-level students help teach those at lower levels \cite{mezmur-tsegaye-tourist:2011}. Additionally, while a student at one level is practicing what the teacher has taught, the teacher can simultaneously instruct students at different levels, one by one. This dual system of peer teaching and simultaneous multilevel instruction is common in all EOTC spiritual schools. However, the centers of excellence level schools, known as \emph{Masmeskeriya} or \emph{Misikir Guba'e Bet} (meaning the schools for the final certification), primarily use the second method, as their students are already educated and are working towards certification \cite{berhanu-akal:2012}.


Before the introduction of modern education in the early 20\textsuperscript{th} century, the EOTC schools served as primary sources of literacy \cite{Molla-pedagogy-eotc:2024, mezmur-tsegaye-tourist:2011}. Although the exact date of inception of these spiritual schools remains unclear \cite{Molla-pedagogy-eotc:2024}, it is evident that they were established before the 6\textsuperscript{th}  century. This is supported by numerous sources indicating that the 6\textsuperscript{th} century marks the rise of Saint Yared, the renowned composer of the EOTC chants (\emph{Yaredawi Zema}) \cite{Kebede:80, GirmaZemedu:Hymn-Synthesis, mezmur-tsegaye-tourist:2011, bekerie2007st, Baye2023:aquaquamZema}. 


\subsection{Modes of Chanting - Yaredawi YeZema Siltoch}

Saint Yared composed five chanting books with the three chanting modes \cite{bekerie2007st}, known as \emph{Yaredawi YeZema Siltoch} in Amharic, namely Ge'ez, Ezil, and Araray. Each mode defines the pitch and stylistic elements of the chants and carries deep symbolic meanings connected to the Holy Trinity \cite{berhanu-akal:2012}. 
Ge'ez, the simplest chant, used on ordinary days \cite{mezmur-tsegaye-tourist:2011}, symbolizes the Father. It is characterized by a hard and imposing style \cite{bekerie2007st} and a low octave range, conveying depth, stability, and solemnity \cite{Kebede:80}.
Ezil, representing the Son, is a melodic and gentle mode employed during fasts and funerals \cite{mezmur-tsegaye-tourist:2011}. It offers a slow, dignified tone with an affective, tender quality \cite{bekerie2007st}, occupying the middle range to provide harmony and neutrality \cite{Kebede:80}. Araray, associated with the Holy Spirit, is the most complex and ornamented mode, used during great festivals and celebrations \cite{mezmur-tsegaye-tourist:2011,Kebede:80}. It features a lively and embellished performance style, conveying joy and cheerfulness, but can also add a melancholic yet melodious note during somber moments like fasting and funerals \cite{bekerie2007st}. These three modes encapsulate the musical and emotional spectrum within the EOTC chants, guiding the emotional content and enriching the spiritual experience through their symbolic and liturgical functions. In light of the significance of these chanting modes, our dataset includes 
such 
labels for each audio.

\subsection{Notations}

\begin{figure}[t]    
\centerline{\includegraphics [width=\columnwidth]{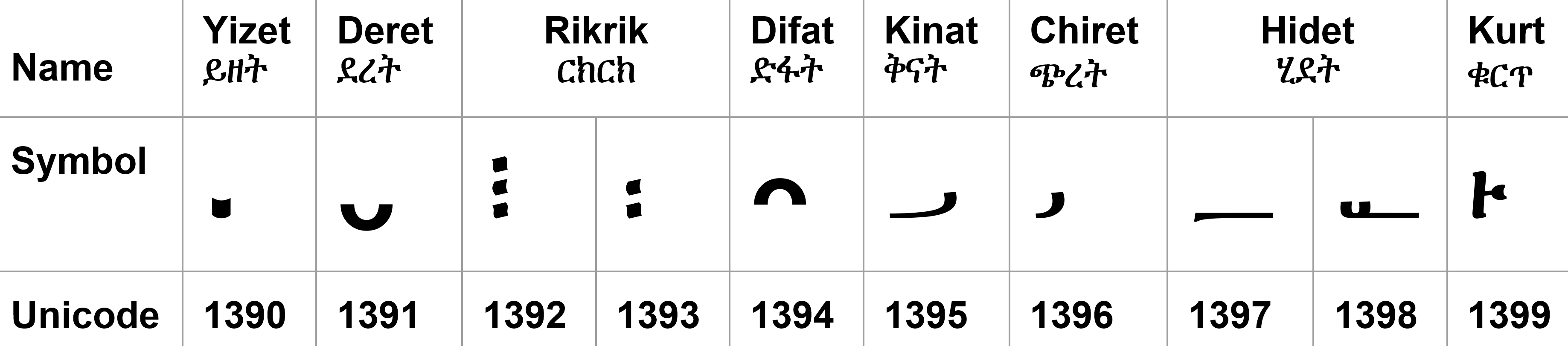}}
     \caption{The eight basic symbolic notations (neumes) in the EOTC chants. For example: \emph{Yizet} represents a detached and accented tone (equivalent to \emph{staccato}), to emphasize a letter or a word. Interested readers on more details are advised to refer to previous articles like \cite{Kebede:80, bekerie2007st}.} 
     \label{fig:symbolic-notations} 
\end{figure}

The EOTC chants feature distinct sets of notations, comprising eight neumes \cite{bekerie2007st}, as depicted in Fig. \ref{fig:symbolic-notations}\footnote{We do not consider Figures \ref{fig:symbolic-notations} and \ref{fig:annotation} as tables because we do not construct them in \LaTeX, rather we easily imported them as screenshot figures of tables containing characters in Ge'ez language.}, and alphabetical notations, termed \emph{miliktoch} (signs) and \emph{sireyu} (basic roots) \cite{Kebede:80}. The \emph{miliktoch} indicate the relative pitch, duration, volume, and performance techniques of each melodic passage. These notations play a crucial role in the tradition of \emph{Yaredawi Zema}, transmitting performance approaches such as tempo, dynamics, and special ornamentation \cite{Kebede:80}. They also convey the melody of singing, making cantillation easier for students. Given their importance, students are encouraged to prioritize mastering these notations, as it facilitates accelerated learning of the subsequent notated content.

\begin{figure}[t]    
\centerline{\includegraphics [width=\columnwidth]{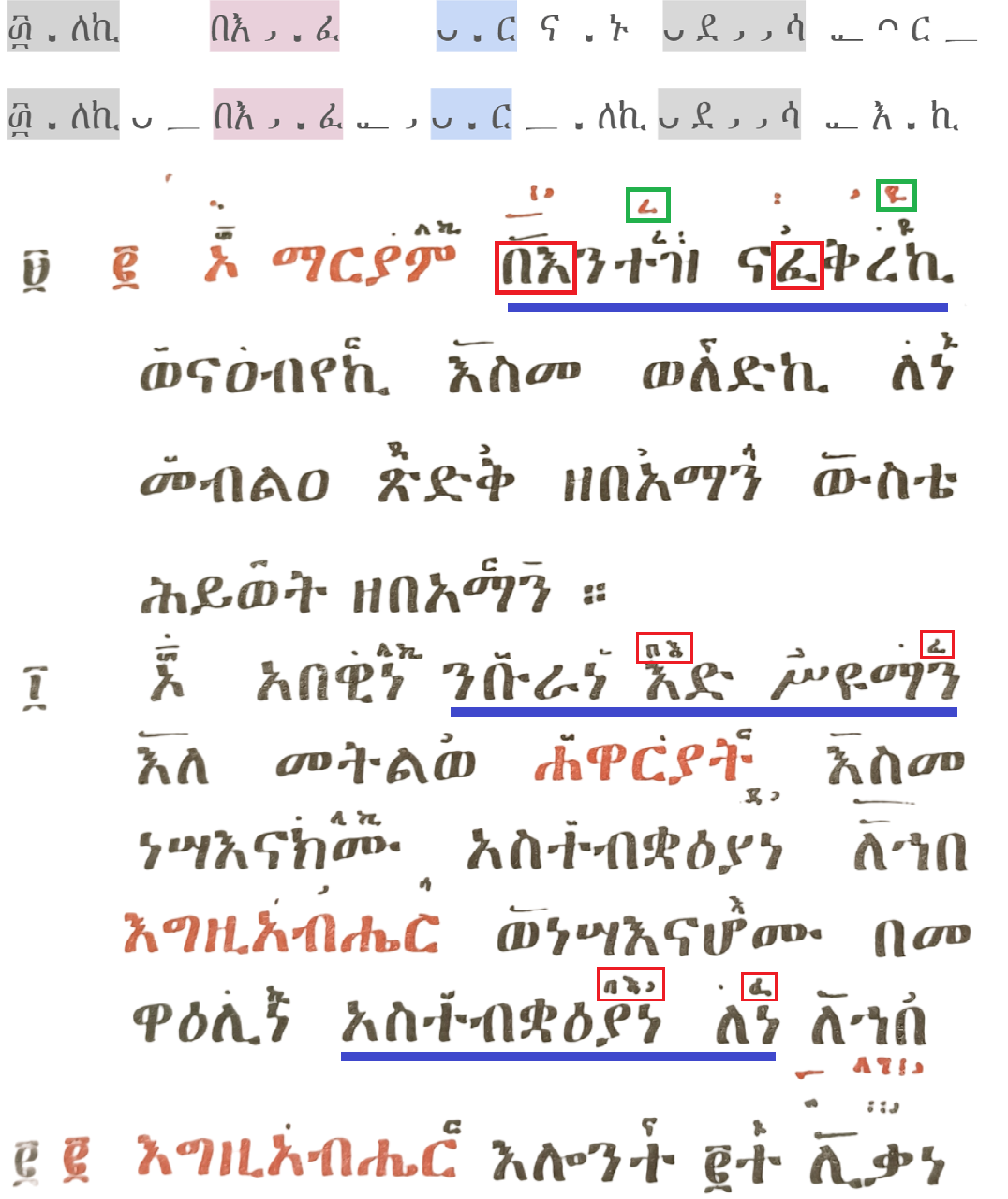}}
     \caption{
     \emph{Sireyu} combined with low-level \emph{milktoch}.
In the 
EOTC chant manuscript, lyrics are written in large texts while music notations are written in small texts above the lyrics. The first two rows of notation patterns above the screenshot are extracted from the two sentences, marked with Ge'ez numbers on their left. The first underlined words represent the \emph{sireyu} while the second and third are immediate example applications of this \emph{sireyu}. While \emph{sireyu} has the two letter notations, encircled in green, indicating its ``\emph{sireyu}ness,'' it is represented by its red-circled short-form characters when applied the same melodies at other places. 
} 
     \label{fig:letter-notations} 
\end{figure}

The use of alphabetical notations, \emph{sireyu}, as illustrated in Fig. \ref{fig:letter-notations}, aids in remembering complex combinations of neumes once the original example is comprehended. Initially, grasping these intricate combinations can be challenging, but with repeated exposure, rehearsing similar instances becomes simpler. For instance, one occurrence in our dataset, where a two-character word is chanted for 45.4 seconds, showcases the multitude of complex combinations within such a structure. A manuscript screenshot of chants \cite{mezgeb-kidase:2012} in Fig. \ref{fig:letter-notations} demonstrates how the \emph{sireyu} are applied to other similarly chanted phrases. The lyrics along each line, written in larger characters, are notated by a combination of \emph{miliktoch} and \emph{sireyu}, facilitating the retrieval of their corresponding melody patterns.


Fig. \ref{fig:letter-notations} presents crucial information on the \emph{sireyu} notation. The first case illustrates the re-usability of a melody for different phrases. In the second case, when applying this \emph{sireyu} to similarly chanted phrases, the accompanying neumes may vary. These variations are determined by the reading tone, 
the number of voiced syllables, 
and the chanting mode. 

A significant amount of the notations, either neumes or sireyu, could be used in two or three of the modes. The same notation thus will have a different melody according to the mode associated to it. Hence, most of the notations will not help identify the mode, except their color. Notations written in red always indicate \emph{Ezil YeZema Silt}, while the black notation mostly represents both \emph{Araray} and \emph{Ge'ez YeZema Siltoch}. 
On the other hand, the black notations in Fig. \ref{fig:letter-notations}, having no red notations above them as options, are common for both \emph{Araray} and \emph{Ezil}.



The third salient piece of information, from Fig. \ref{fig:letter-notations}, is the general recurring melody patterns in chanting sentences. The two lines, above the screenshot, demonstrate the relationship between the chanting patterns of the two sentences. A similar melodic flow is consistent throughout the book. Although not all sentences follow the same pattern, there are a certain number of patterns that can be identified from the beginning to the end of sentences. Lastly, it is worth noting that these symbols are used in the same book or department and also in chanting within other departments, possibly with some ornamental customization based on the category. 

The notation system of \emph{Yaredawi Zema} has a profound cultural and religious connection, having been passed down through generations of highly skilled musicians \cite{shelemay1993oral}. This deeply rooted tradition is not limited to spiritual contexts; its applicability in non-spiritual written texts is claimed in \cite{Kebede:80}. Moreover, it has been demonstrated to have the capacity to prepare melodies for any written text in any language \cite{mezmur-tsegaye-tourist:2011}.



\subsection{Chanting Options} 

\begin{figure*}[t]
    \centering
    \begin{subfigure}{0.52\textwidth}
        \includegraphics[width=\linewidth]{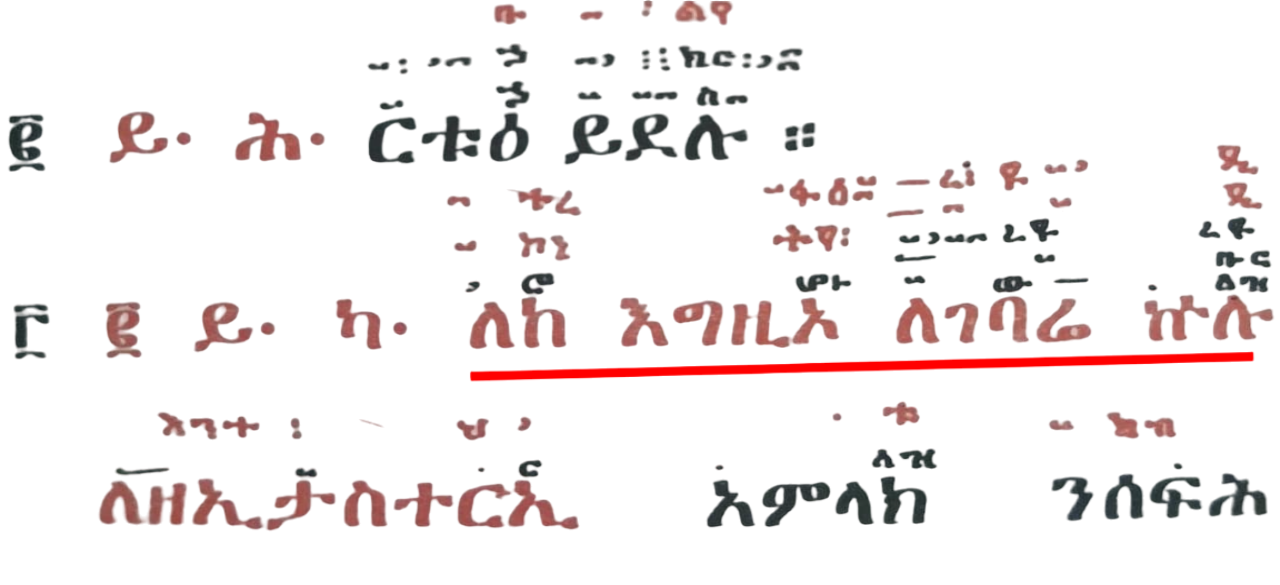}
        \caption{Multiple ways of chanting within each mode, illustrated by the underlined four-word phrase with two chanting options in Ezil mode (red notations) and three chanting options in Ge'ez mode (black notations).}
        \label{fig:optional-notations}
    \end{subfigure}\hfill
    \begin{subfigure}{0.44\textwidth}
        \includegraphics[width=\linewidth]{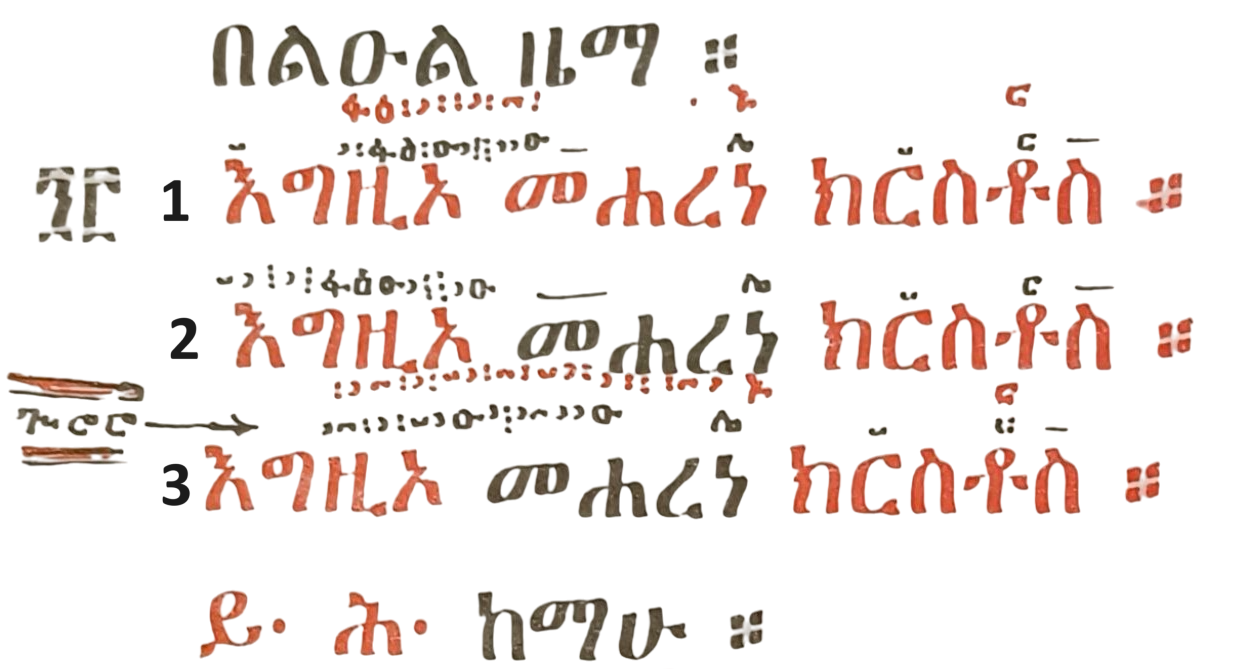}
        \caption{Non-optional repeated phrases in both the audio and book, not chanting options despite seeming so during annotation without referencing the book.} 
        \label{fig:non-optional-repetitions}
    \end{subfigure}
    \caption{
    Illustration examples of identifying chanting options of repeatedly appearing phrases in audios, each occurrence with different notations. Identification needs some background knowledge and referring to notated manuscripts, applicable in all chanting modes, shown here for Ezil and Ge'ez. }
    \label{fig:chanting-options}
\end{figure*}

We identified two types of chanting options in the audio records: optional words with the same melodic notation and multiple notations placed above the lyrics text, as shown in Fig. \ref{fig:optional-notations}. This study focuses on annotating the latter type. 

As illustrated in Fig. \ref{fig:optional-notations}, the three-line screenshot of a notated manuscript showcases chanting in both \emph{Ge'ez} and \emph{Ezil} modes, where the choice of mode usually depends on the season and whether it is a holiday or fasting period. Each chanting mode encompasses a significant number of words and phrases that can be chanted in multiple ways. Considering all available options, a single word or phrase can be chanted in various ways. For example, the underlined phrase, when chanted in the \emph{Ezil} mode, offers two chanting options indicated by the red notations. Conversely, in \emph{Ge'ez} mode, the last two words of the underlined phrase present three chanting options, marked by black notations. Note that such phrases having optional chanting notations appear once on the book of chant with all multiple notations placed above them. However, as discussed in Section \ref{subsubsec:text-preproc}, we duplicated them according to the number of options included in the recordings, the above-mentioned three options of \emph{Ge'ez} mode are repeated three times as shown in Fig. \ref{fig:annotation}.  

Recognizing the chanting options aided by notated manuscripts and the corresponding audio is typically straightforward. However, it becomes challenging when the audio is accompanied by text that contains repeated phrases for different options. This ambiguity arises because sometimes a repeated phrase may have a different notation combination than its previous occurrence, but the repeated occurrences may not represent an alternative way of chanting. In such cases, consecutively repeated phrases in the recordings also appeared repeatedly in the notated manuscript, as shown in Fig. \ref{fig:non-optional-repetitions}. In the annotation stages, this may lead to confusion regarding whether we repeated a phrase for its multiple notations or if the book repeated the phrase as a non-optional way. 


\section{Dataset}\label{sec:dataset} 
\subsection{Data Source}\label{subsec:datasource}
The audio dataset was manually collected from the \emph{Eat the Book} website\footnote{\url{https://eathebook.org/}, We sincerely appreciate the website administrators for making a wealth of learning resources publicly available.}, a hub of numerous audiobooks for most of the teachings of the EOTC schools. From the audiobooks available for these schools, we selected the \emph{Se'atat Zema} (Horologium chant), which is part of the \emph{Qidase-bet} department. All audios selected for our dataset were recorded by a single scholar at a sampling rate of 44,100 Hz and in stereo channel.

\subsection{Pre-processing}\label{subsec:preprocessing}

A significant amount of effort is invested in cleaning and organizing both the audio recordings and the textual scripts. The preprocessing of data includes the following steps. 

\subsubsection{Audio Cleaning and Arrangement} \label{subsubsec:audio-preproc}

To facilitate data processing, our initial step in pre-processing the audio recordings involved reducing the duration disparity between long and short audio recordings. Long audios exceeding 13 minutes were segmented into shorter clips of less than three minutes (180 seconds) while preserving their full context. This segmentation was also applied to audios that contain multiple chanting modes. For instance, audio with 160 seconds of the \emph{Araray} mode followed by 22 seconds of \emph{Ezil} mode was divided into two separate clips of 160 seconds and 22 seconds, respectively. Even recordings shorter than three minutes but containing multiple modes were segmented according to the duration of each chanting mode.

Conversely, short audios, such as those lasting only 19 seconds, were merged with adjacent audios of the same mode, when applicable. If no neighboring audio with the same mode was available, the short audio was treated as a separate clip. In short, each audio was arranged to contain a single mode, resulting in a corresponding mode label for each clip. Non-chant segments, including explanatory statements about the chants, were manually removed to ensure the audio content matched the lyrics. After these cleaning procedures, the overall duration distribution of our dataset ranged from 20.142 seconds to 177.476 seconds. After preprocessing, our dataset then contains more than 10 hours and 369 instances.

\subsubsection{Text Extraction and Preprocessing}\label{subsubsec:text-preproc}


Text corresponding to the audio recordings was obtained from two sources: 1) a PDF document containing Ge'ez chants (text extracted using a Python script),\footnote{Seatat for church service\_august2016.pdf, a presentation released in a PDF format on \url{https://www.ethiopianorthodox.org}, Ge'ez texts colored yellow whereas Amharic texts colored white.} 
and 2) photographs of the book's  \cite{yeneta-yemane:2010} pages processed with Google Lens. 
Data cleaning is required for both sources, such as the removal of Amharic translations in 1) and some misinterpreted texts in 2). 



\begin{figure}[t]
    \centerline{\includegraphics[width=0.95\columnwidth]{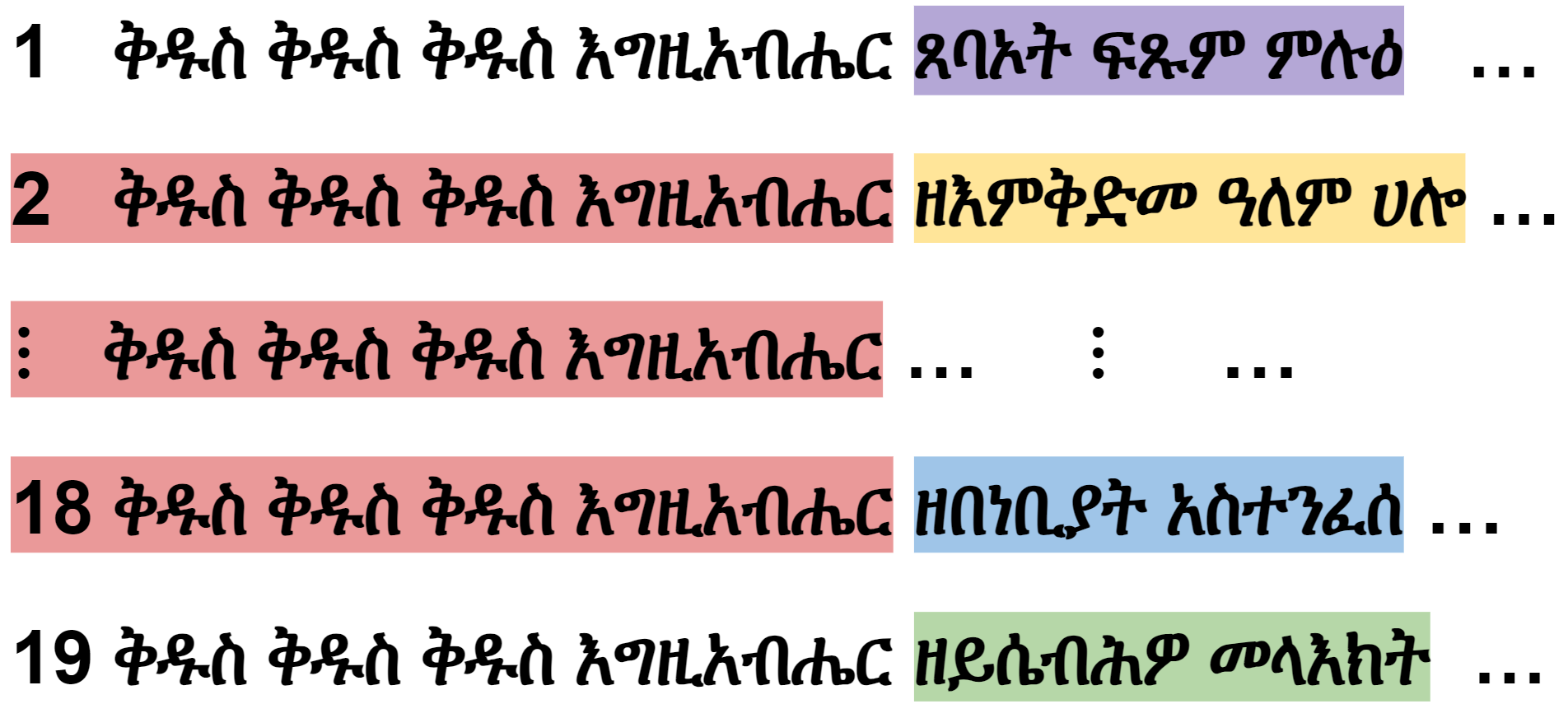}}
    \caption{Illustration of repeated phrases with identical notations in the chant text. Lines 2 to 18 have their first four words omitted due to sharing the same melody as in line 1, while the remaining words vary and are retained. The first four words in line 19 are chanted with the same melody as in line 1, serving as a reference before concluding this pattern of repetition. Similar cases are found throughout our dataset. 
    } 
    \label{fig:repeated-text} 
\end{figure}

\begin{figure*}[htbp]
    \centerline{\includegraphics[width=2\columnwidth]{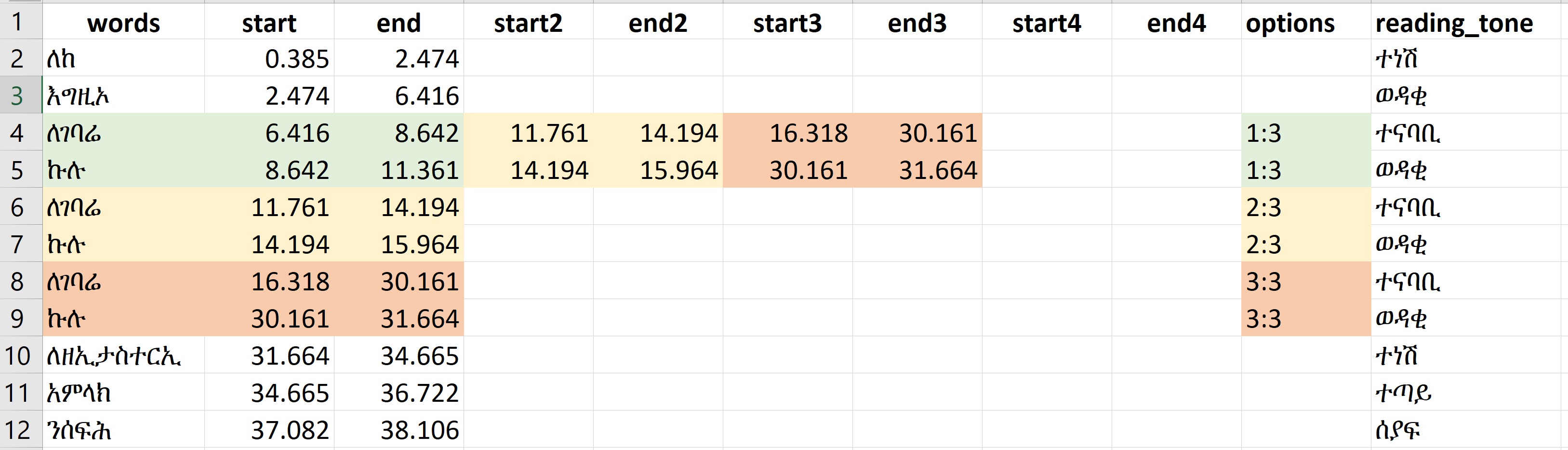}}
    \caption{Sample annotation, for \emph{Ge'ez YeZema silt} of the text shown in Fig. \ref{fig:optional-notations}. As there are three chanting options in this mode, we duplicated the two-word phrase three times, though the text appeared once in the book of chant. If we consider the annotation of a recording in \emph{Ezil YeZema silt} of the same text, the underlined phrase will be duplicated twice as explained in Section \ref{subsubsec:text-preproc}. Regarding annotation columns of start4 and end4, for a fourth chanting option in the same mode, we found only two cases and empty values in all others. }
    \label{fig:annotation} 
\end{figure*}

 
A common feature in the EOTC chants we considered is the repetition of phrases in consecutive lines of text, as shown in Fig. \ref{fig:repeated-text}, which have identical notation. In educational recordings, these repeated phrases are usually only included in their first and last occurrence, with the middle repetitions omitted. Therefore, these repeated texts excluded from the recordings need to be removed to achieve an exact audio-lyric match. In addition, text was duplicated for chants with optional variations in the same mode, as illustrated in Figure \ref{fig:annotation}, to ensure that all audio variations were represented. 
In essence, text preprocessing involved extraction, removal of unwanted elements, handling repetitive text and optional chants, spelling correction, tokenization (splitting into words), and the assignment of annotation headings. 

\subsection{Data Annotation Process} \label{subsec:annotation-process} 

In this study, the annotation process involves determining the precise start and end temporal boundaries of words in the recordings using Audacity software\footnote{\url{https://www.audacityteam.org/}}. 
To accomplish this, we open the audio files in Audacity and zoom in as much as possible to accurately identify the temporal boundaries of each word. By carefully listening and following the progress of the audio signal in Audacity, we pinpoint the specific segment of the signal that corresponds to each word. The start and end times 
(in seconds) 
of the identified audio segments are then recorded in the corresponding Excel file for the audio. The start and end time values denote the start and end times of each word, taking into account any optional occurrences.

To systematically document the identified temporal boundaries, the file names of the audio and the label are identical, aside from the file extension. For example, if we open an audio file named "sample.mp3" in Audacity, we concurrently open the corresponding "sample.xlsx" Excel file to record the start and end time of each word chanted in "sample.mp3". 
As shown in Fig. \ref{fig:annotation}, the end time of a word may coincide with the start time of the subsequent word, although it is common for the start time of the subsequent word to be different from the end time of the previous word. Furthermore, to clearly represent chanting options, "start" and "end" time values of repeated occurrences are duplicated as "start2, end2" and "start3, end3" values of the first occurrence, and the corresponding optional labels are recorded in the "options" column. 
The options label consists of two numbers separated by a colon, indicating the number of options available. For example, if the "options" label gets a value of \emph{1:3}, it means that the word in the specified time gap is chanted using the first option out of the three available options. Finally, reading tone annotations were included as well, since they convey important information about the notation combinations. To the best of our knowledge, we prepared the largest feature-rich dataset for the case of EOTC chants, which is about 10.21 hours of data.



\begin{figure*} 
\centerline{\includegraphics [width=\textwidth]{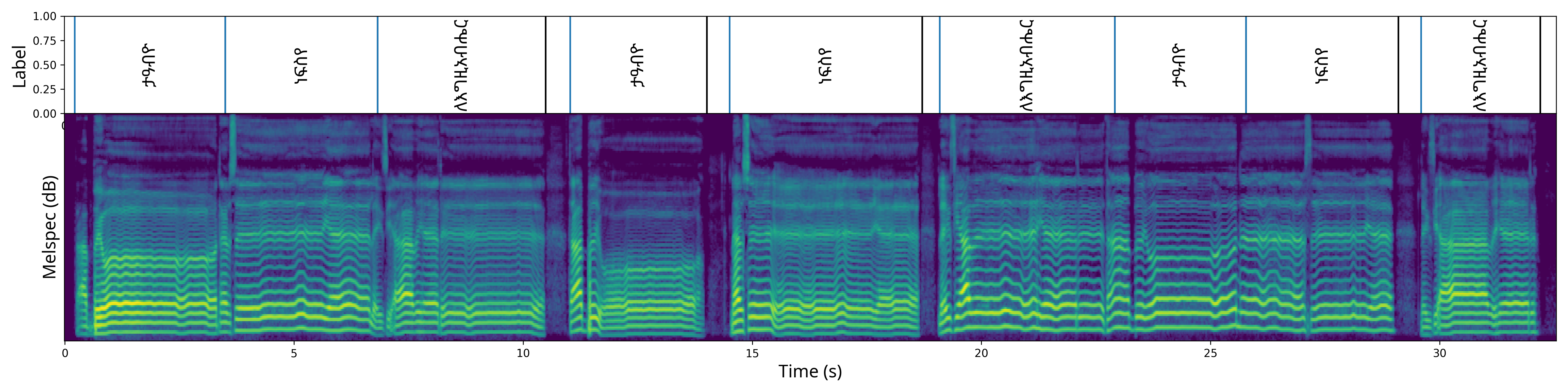}} 
     \caption{Melspectrogram 
     (lower part) 
     aligned with words 
     (upper part)
     as a high-level method of annotation quality inspection. 
     For the world labels, a blue vertical line represents the start time of a word, while a black vertical line represents the end time.} 
     \label{fig:melspec-inspection}
\end{figure*}

To ensure data quality, several validation steps were conducted: handling missing values, validating data types, ensuring temporal value consistency, and correcting mislabeled chanting options due to non-optional repetitions (Fig. \ref{fig:non-optional-repetitions}). Furthermore, we performed a visual assessment of annotation quality by plotting the melspectrogram of each audio file alongside the word-label annotations (Fig. \ref{fig:melspec-inspection}).

\subsection{Annotators}

Three annotators were invited to annotate part of  the data 
accounting for 6 hours. 
In addition, a certified expert in Ge'ez language was invited to correctly annotate the reading tone of all words in the full 10.21 hours of dataset. As some dataset papers like \cite{bps:2023} included reviews, we also collected a self-evaluation form 
to reflect the three annotators' expertise with regard to the given task. Our questionnaire included the following questions. 

\begin{itemize}
    \item \textbf{Q1) }How do you describe your level of awareness of the Ge’ez language and the EOTC chants?  4) Excellent 
3) Good 
2) Satisfactory 
1) Poor 
    \item \textbf{Q2)} How confident are you in the accuracy of your annotations when capturing the timings of words in the lyrics of chants? 4) Totally accurate 3) Mostly accurate 2) Accurate 1) Poor 
    \item \textbf{Q3)} Did you strictly keep coherence to the annotation guidelines provided? If not, what deviations did you make and why?  4) Totally coherent 3) Mostly coherent 2) Partially coherent 1) Not coherent  
    \item \textbf{Q4)} Did you ensure consistency in your timing annotations across different aspects? If not, what factors contributed to inconsistencies? 4) Totally consistent 3) Mostly consistent 2) Consistent 1) Poor 
    \item \textbf{Q5)} How effectively did you handle chanting options? 4) Handled effectively  3) Managed adequately 2)  Struggled 1) Unable to handle 
    \item \textbf{Q6)} Did you consider reviewing your work to ensure the quality of your annotations? a) Rigorous review b)  Reviewed c) Limited review d) No review
\end{itemize}


\begin{figure} 
    \centering    \includegraphics[width=\linewidth]{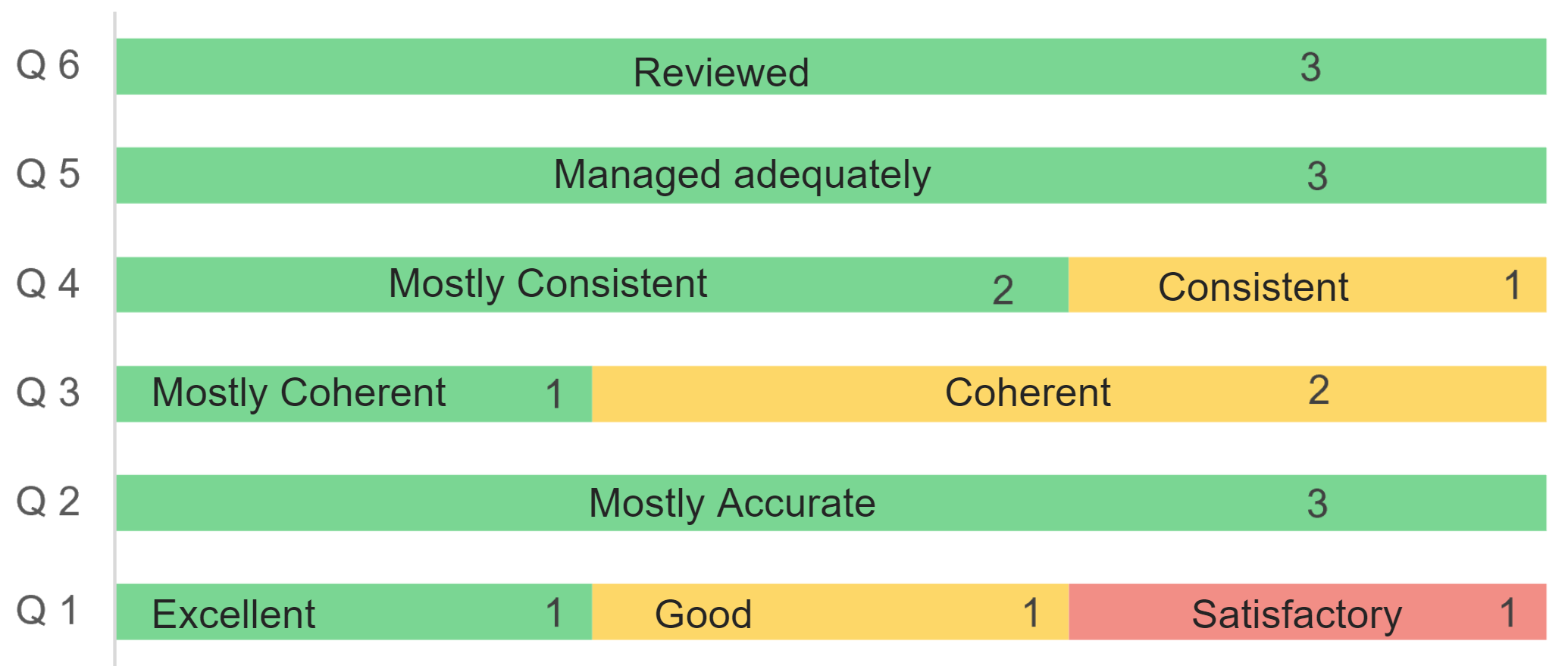}
    \caption{Self-evaluation (Q1 to Q6) of the annotators.}
    \label{fig:questionnaire}
\end{figure}

The results of our self-evaluation questionnaire are shown in Fig. \ref{fig:questionnaire}. 
Our annotators possess a satisfactory to excellent understanding of the Ge'ez language and the chants, sufficient for recognizing word-level timings and chanting options. All annotators accurately captured timing labels, adhered closely to the provided guidelines, and demonstrated mostly consistent timing expectations. In addition, they effectively captured the chanting options and conducted cross-reviews of their annotations. The three annotators reported approximately 200 hours for the first-round annotations and nearly 100 hours for quality double-checking. 


 

\subsection{Case study}

We present a preliminary case study using the proposed dataset for the classification of the chanting mode of \emph{Yaredawi Zema}. We employed a Support Vector Machine (SVM) with an RBF kernel (C=1, gamma='scale') for classifying chanting modes using mel frequency cepstral coefficient (MFCC) features. Utilizing a StandardScaler for feature standardization, we evaluated the model using 10-fold cross-validation, resulting in an overall accuracy of 69.11\%. The confusion matrix, shown in Table \ref{tab:confusion-matrix}, highlights the distribution of correct and incorrect predictions across three chanting modes: Araray, Ezil, and Ge'ez. The Ezil chanting mode demonstrates best recognition rate (i.e., 85\%) due to its distinct nature and higher pitch range. Conversely, the Ge'ez mode is relatively harder to be classified, probably because of its lower pitch ranges. 

\begin{table}[t]
\caption{Confusion matrix on a case study of chanting mode recognition using SVM with 10-fold cross-validation.}
 \begin{center}
 \begin{tabular}{|c|ccc|} 
 \hline
   & Array & Ezil& Ge'ez \\\hline
 Array & 0.62& 0.26& 0.12\\
 Ezil & 0.12& 0.85& 0.03\\
 Ge'ez & 0.27 &0.30& 0.43\\  \hline  
 \end{tabular}
\end{center}
 
 \label{tab:confusion-matrix}
\end{table}

\section{Conclusion}\label{sec:conclusion}
This paper presents a meticulously organized, cleaned, and annotated dataset specifically designed for various music information retrieval and music generation tasks, focusing on the under-explored Ethiopian Orthodox Tewahedo Church (EOTC) chant. Our dataset concentrates on the \emph{Se'atat Zema} (Horologium chant) from the \emph{Qidasie-bet} school, aiming to encourage responsible research on the EOTC chants and foster technological advancements that enhance learning and increase accessibility.

 

The feature-rich dataset includes detailed annotations such as word-level temporal boundaries, chanting options, reading tones, and chanting mode categories for each recording. These features enable the dataset to support a variety of deep learning tasks, including chanting mode (\emph{YeZema Silt}) recognition, lyrics transcription, lyrics-to-audio alignment, and music generation, all customized to \emph{Yaredawi Zema}. 
Additionally, it supports multimodal and various music analysis tasks, contributing to a deeper understanding of the chants. 

\section{Acknowledgment} 
We would like to thank Mr. Dagne Almaw Kassa for his support in annotating the reading tones of all words in our dataset. 
\bibliographystyle{IEEEtran}
\bibliography{ICT4DA_2024}

\begin{thebibliography}{10}
\providecommand{\url}[1]{#1}
\csname url@samestyle\endcsname
\providecommand{\newblock}{\relax}
\providecommand{\bibinfo}[2]{#2}
\providecommand{\BIBentrySTDinterwordspacing}{\spaceskip=0pt\relax}
\providecommand{\BIBentryALTinterwordstretchfactor}{4}
\providecommand{\BIBentryALTinterwordspacing}{\spaceskip=\fontdimen2\font plus
\BIBentryALTinterwordstretchfactor\fontdimen3\font minus \fontdimen4\font\relax}
\providecommand{\BIBforeignlanguage}[2]{{%
\expandafter\ifx\csname l@#1\endcsname\relax
\typeout{** WARNING: IEEEtran.bst: No hyphenation pattern has been}%
\typeout{** loaded for the language `#1'. Using the pattern for}%
\typeout{** the default language instead.}%
\else
\language=\csname l@#1\endcsname
\fi
#2}}
\providecommand{\BIBdecl}{\relax}
\BIBdecl

\bibitem{berhanu-akal:2012}
\BIBentryALTinterwordspacing
B.~A. Abebe, ``Content analysis of matshafa mawas’et,'' Master's Thesis, Addia Ababa University, Addis Ababa, Ethiopia, 2012. [Online]. Available: \url{http://etd.aau.edu.et/handle/123456789/248}
\BIBentrySTDinterwordspacing

\bibitem{Kebede:80}
A.~Kebede, ``The sacred chant of {E}thiopian monotheistic churches: Music in black {J}ewish and {C}hristian communities,'' in \emph{The Black Perspective in Music}, J.~Southern, Ed.\hskip 1em plus 0.5em minus 0.4em\relax Brandeis University: JSTOR, 1980, pp. 20--34.

\bibitem{mezmur-tsegaye-tourist:2011}
\BIBentryALTinterwordspacing
M.~Tsegaye, ``Traditional education of the ethiopian orthodox church and its potential for tourism development (1975-present),'' Master's Thesis, Addia Ababa University, Addis Ababa, Ethiopia, 2011. [Online]. Available: \url{http://etd.aau.edu.et/handle/123456789/248}
\BIBentrySTDinterwordspacing

\bibitem{PIKHART-ai4language-2020}
\BIBentryALTinterwordspacing
M.~Pikhart, ``Intelligent information processing for language education: The use of artificial intelligence in language learning apps,'' \emph{Procedia Computer Science}, vol. 176, pp. 1412--1419, 2020, knowledge-Based and Intelligent Information \& Engineering Systems: Proceedings of the 24th International Conference KES2020. [Online]. Available: \url{https://www.sciencedirect.com/science/article/pii/S1877050920320512}
\BIBentrySTDinterwordspacing

\bibitem{ZHANG-AI-Education:2021}
\BIBentryALTinterwordspacing
K.~Zhang and A.~B. Aslan, ``Ai technologies for education: Recent research \& future directions,'' \emph{Computers and Education: Artificial Intelligence}, vol.~2, p. 100025, 2021. [Online]. Available: \url{https://www.sciencedirect.com/science/article/pii/S2666920X21000199}
\BIBentrySTDinterwordspacing

\bibitem{youtube-recsys:2016}
\BIBentryALTinterwordspacing
P.~Covington, J.~Adams, and E.~Sargin, ``Deep neural networks for youtube recommendations,'' in \emph{Proceedings of the 10th ACM Conference on Recommender Systems}, ser. RecSys '16.\hskip 1em plus 0.5em minus 0.4em\relax New York, NY, USA: Association for Computing Machinery, 2016, p. 191–198. [Online]. Available: \url{https://doi.org/10.1145/2959100.2959190}
\BIBentrySTDinterwordspacing

\bibitem{netflix-recsys:2016}
\BIBentryALTinterwordspacing
C.~A. Gomez-Uribe and N.~Hunt, ``The netflix recommender system: Algorithms, business value, and innovation,'' \emph{ACM Trans. Manage. Inf. Syst.}, vol.~6, no.~4, dec 2016. [Online]. Available: \url{https://doi.org/10.1145/2843948}
\BIBentrySTDinterwordspacing

\bibitem{han2023finding}
D.~Han, R.~C. Repetto, and D.~Jeong, ``Finding tori: Self-supervised learning for analyzing korean folk song,'' in \emph{International Society of Music Information Retrieval Conference (ISMIR)}, 2023.

\bibitem{rosenzweig2020erkomaishvili}
S.~Rosenzweig, F.~Scherbaum, D.~Shugliashvili, V.~Arifi-M{\"u}ller, and M.~M{\"u}ller, ``Erkomaishvili dataset: A curated corpus of traditional {G}eorgian vocal music for computational musicology.'' \emph{Transactions of International Society of Music Information Retrieval}, vol.~3, no.~1, pp. 31--41, 2020.

\bibitem{nikzat2022kdc}
B.~Nikzat and R.~Caro~Repetto, ``{KDC}: an open corpus for computational research of dastg{\=a}hi music,'' in \emph{International Society of Music Information Retrieval Conference (ISMIR)}, 2022.

\bibitem{caro2014creating}
R.~Caro~Repetto and X.~Serra, ``Creating a corpus of jingju (beijing opera) music and possibilities for melodic analysis,'' in \emph{International Society of Music Information Retrieval Conference (ISMIR)}, 2014.

\bibitem{nadkarni2023exploring}
S.~Nadkarni, S.~Roychowdhury, P.~Rao, and M.~Clayton, ``Exploring the correspondence of melodic contour with gesture in raga alap singing,'' in \emph{International Society of Music Information Retrieval Conference (ISMIR)}, 2023.

\bibitem{lisu:amt-sitar-music}
\BIBentryALTinterwordspacing
A.~C. Li~Su and Y.-F. Huang, ``Automatic music transcription for sitar music analysis,'' \emph{Journal of New Music Research}, vol.~51, no. 4-5, pp. 278--299, 2022. [Online]. Available: \url{https://doi.org/10.1080/09298215.2023.2251450}
\BIBentrySTDinterwordspacing

\bibitem{moss2023line}
F.~C. Moss, M.~Neuwirth, and M.~Rohrmeier, ``The line of fifths and the co-evolution of tonal pitch-classes,'' \emph{Journal of Mathematics and Music}, vol.~17, no.~2, pp. 173--197, 2023.

\bibitem{lieck2020tonal}
R.~Lieck, F.~C. Moss, and M.~Rohrmeier, ``The tonal diffusion model.'' \emph{Transactions of the International Society for Music Information Retrieval}, vol.~3, no.~1, p. 153, 2020.

\bibitem{music-culture:2021}
\emph{The Role of Music Education in Cultural Preservation, Perpetuation and Development in 21st Century Digital Environments}, 2021.

\bibitem{weiss2021schubert}
C.~Wei{\ss}, F.~Zalkow, V.~Arifi-M{\"u}ller, M.~M{\"u}ller, H.~V. Koops, A.~Volk, and H.~G. Grohganz, ``Schubert winterreise dataset: A multimodal scenario for music analysis,'' \emph{Journal on Computing and Cultural Heritage (JOCCH)}, vol.~14, no.~2, pp. 1--18, 2021.

\bibitem{stepputat2019digital}
\BIBentryALTinterwordspacing
K.~Stepputat, W.~Kienreich, and C.~S. Dick, ``Digital methods in intangible cultural heritage research: A case study in tango argentino,'' \emph{ACM Journal on Computing and Cultural Heritage}, vol.~12, no.~2, may 2019. [Online]. Available: \url{https://doi.org/10.1145/3279951}
\BIBentrySTDinterwordspacing

\bibitem{rosenzweig2022computer}
\BIBentryALTinterwordspacing
S.~Rosenzweig, F.~Scherbaum, and M.~M\"{u}ller, ``Computer-assisted analysis of field recordings: A case study of {G}eorgian funeral songs,'' \emph{ACM Journal on Computing and Cultural Heritage}, vol.~16, no.~1, dec 2022. [Online]. Available: \url{https://doi.org/10.1145/3551645}
\BIBentrySTDinterwordspacing

\bibitem{holzapfel2018ethical}
A.~Holzapfel, B.~Sturm, and M.~Coeckelbergh, ``Ethical dimensions of music information retrieval technology,'' \emph{Transactions of the International Society for Music Information Retrieval}, vol.~1, no.~1, pp. 44--55, 2018.

\bibitem{Dagne-eotc-org}
\BIBentryALTinterwordspacing
H.~G. Dagne, ``The ethiopian orthodox church school system,'' accessed on May 31, 2024. [Online]. Available: \url{https://www.ethiopianorthodox.org/english/ethiopian/school.html}
\BIBentrySTDinterwordspacing

\bibitem{Molla-pedagogy-eotc:2024}
\BIBentryALTinterwordspacing
A.~B.~T. Molla Bekalu~Mulualem and K.~A. Dagnew, ``The pedagogical practices of ethiopian orthodox church traditional schools: Implications for contemporary education,'' \emph{Pedagogy, Culture \& Society}, vol.~32, no.~1, pp. 257--273, 2024. [Online]. Available: \url{https://doi.org/10.1080/14681366.2022.2027808}
\BIBentrySTDinterwordspacing

\bibitem{GirmaZemedu:Hymn-Synthesis}
G.~Zemedu and Y.~Assabie, ``Concatenative hymn synthesis from {Y}ared notations,'' in \emph{Advances in Natural Language Processing}, A.~Przepi{\'o}rkowski and M.~Ogrodniczuk, Eds.\hskip 1em plus 0.5em minus 0.4em\relax Cham: Springer International Publishing, 2014, pp. 400--411.

\bibitem{bekerie2007st}
\BIBentryALTinterwordspacing
A.~Bekerie, ``St. {Y}ared--the great {E}thiopian composer,'' \emph{Tadias Magazine, New York}, 2007. [Online]. Available: \url{http://www.tadias.com/11/29/2007/st-yared-the-great-ethiopian-composer/}
\BIBentrySTDinterwordspacing

\bibitem{Baye2023:aquaquamZema}
B.~T. Dagnew, ``Ethiopian {O}rthodox {T}ewahido {C}hurch {A}quaquam {Z}ema classification model using deep learning,'' Master's Thesis, Bahir Dar University, Bahir Dar, Ethiopia, 2023.

\bibitem{mezgeb-kidase:2012}
E.~O.~T. Church, \emph{Book of Liturgy in Ge'ez and Amharic with Notations}, revised edition~ed.\hskip 1em plus 0.5em minus 0.4em\relax Addis Ababa, Ethipia: GebreSilassie Berhanu, 2012, origin: Book of Liturgy in Ge'ez and Amharic with Notations.

\bibitem{shelemay1993oral}
K.~K. Shelemay, P.~Jeffery, and I.~Monson, ``Oral and written transmission in {E}thiopian {C}hristian chant,'' \emph{Early Music History}, vol.~12, pp. 55--117, 1993.

\bibitem{yeneta-yemane:2010}
A.~Giyorgis, \emph{Metsihafe Se'atat}.\hskip 1em plus 0.5em minus 0.4em\relax Addis Ababa, Ethipia: Yemaneberhan Getahun, Akotet PLC, 2010.

\bibitem{bps:2023}
Y.-W. Hsiao, T.-P.~C. Tzu-Yun~Hung, and L.~Su, ``Bps-motif: A dataset for repeated pattern discovery of polyphonic symbolic music,'' in \emph{International Society of Music Information Retrieval Conference (ISMIR)}, 2023.

\end{thebibliography}

\end{document}